\documentclass{iau_FM}
\usepackage{graphicx}
\usepackage{natbib}
\usepackage{amsmath}
\usepackage{mathtools}
\usepackage{amssymb}
\usepackage{wasysym}
\usepackage{verbatim}

\newcommand{\aap}{A\&A}
\newcommand{\apj}{ApJ}
\newcommand{\mnras}{MNRAS}

\newcommand{\nat}{Nature}

\newcommand{\apjs}{ApJS}
\newcommand{\apjl}{ApJL}

\newcommand{\rmgal}{{\rm RM_{gal}}}
\newcommand{\rmqso}{{\rm RM_{qso}}}

\newcommand{\rmc}{{\rm RM_{c}}}
\newcommand{\rmt}{{\rm RM_{t}}}

\title[FM 8.~~Statistical properties of RM] 
{Statistical properties of Faraday rotation measure from large-scale magnetic fields in intervening disc galaxies}

\author[A. Basu et al.]   
{Aritra Basu,$^{1,2}$
 S. A. Mao,$^2$ Andrew Fletcher,$^3$ Nissim Kanekar,$^4$ Anvar Shukurov,$^3$
Dominic Schnitzeler,$^2$ Valentina Vacca$^5$
 \and Henrik Junklewitz$^6$}

\affiliation{$^1$Fakult{\"a}t f{\"u}r Physik, Universit{\"a}t Bielefeld, Postfach 100131, 33501 Bielefeld, Germany\\
$^2$Max-Planck-Institut f{\"u}r Radioastronomie, Auf dem H{\"u}gel 69, D-53121 Bonn, Germany\\
email: {\tt aritra@physik.uni-bielefeld.de}\\ [\affilskip]
$^3$School of Mathematics and Statistics, Newcastle University, \\
Newcastle-upon-Tyne, NE13 7RU, UK\\ [\affilskip]
$^4$National Centre for Radio Astrophysics, TIFR, Post Bag 3, Ganeshkhind,\\
 Pune 411007, India\\[\affilskip]
$^5$INAF-Osservatorio Astronomico di Cagliari, Via della Scienza 5,\\
 I-09047 Selargius (CA), Italy\\[\affilskip]
$^6$Argelander Institut f{\"u}r Astronomie, Universit{\"a}t Bonn, Auf dem H{\"u}gel 71, \\
53121 Bonn, Germany}

\pubyear{2015}
\jname{Astronomy in Focus, Volume 1} 
\editors{Piero Benvenuti, ed.}
\begin{document}

\maketitle

\begin{abstract}
To constrain the large-scale magnetic field strengths in cosmologically distant
galaxies, we derive the probability distribution function of Faraday rotation
measure (RM) when random lines of sight pass through a sample of disc galaxies,
with axisymmetric large-scale magnetic fields. We find that the width of the RM
distribution of the galaxy sample is directly related to the mean large-scale
field strength of the galaxy population, provided the dispersion within the
sample is lower than the mean value. In the absence of additional constraints
on parameters describing the magneto-ionic medium of the intervening galaxies,
and in the situation where RMs produced in the intervening galaxies have
already been statistically isolated from other RM contributions along the lines
of sight, our simple model of the magneto-ionic medium in disc galaxies
suggests that the mean large-scale magnetic field of the population can be
measured to within $\sim 50\%$ accuracy.

\end{abstract}

\firstsection 
\section{Introduction}

The cosmic evolution of magnetic fields on scales $\gtrsim1$ kpc remains an
open question in observational astronomy. In order to constrain cosmic
evolution of large-scale magnetic fields in galaxies, it is crucial to measure
their redshift evolution. Due to the faintness of distant galaxies, it is a
challenging proposition to measure magnetic fields directly through their
polarized synchrotron emission. A powerful tool to probe magnetic fields in
distant galaxies is provided by statistical studies of the Faraday rotation
measure (RM) of quasar absorption line systems, which are tracers of galaxies
in the high redshift Universe. The distribution of RM of two quasar samples,
with and without absorption line systems, can be statistically compared to
infer the properties of magnetic fields in the intervening galaxy population
\citep[e.g.,][]{oren95, berne08, joshi13, farne14}. The sample of quasars with
intervening galaxies is referred to as the `target' sample and the sample
without intervening galaxies is called the `control' sample. 

Part of the RM measured towards a quasar which has an intervening galaxy
arises from the quasar itself ($\rmqso$), the intergalactic medium (IGM; $\rm
RM_{IGM}$) and interstellar medium of the Milky Way ($\rm RM_{MW}$).
Thus, the net RM ($\rmt$) measured in the observer's frame for a single line of
sight in the target sample is given by,
\begin{equation}
\rmt = \frac{\rmgal}{(1+z_{\rm gal})^2} + {\mathrm{RM}^\prime_{\rm qso}},
\end{equation}
where $\rmgal$ is the RM contribution from the intervening galaxy and $z_{\rm
gal}$ is its redshift. $\rm RM_{qso}^\prime$ contains RM contributions from
rest of the line of sight and is given by $\rm RM_{qso}^\prime =
{\rmqso}\,{(1+z_{\rm qso})^{-2}} + RM_{IGM} + RM_{MW} + \delta_{\rm RM}$.
Here, $\delta_{\rm RM}$ is the measurement noise and $z_{\rm qso}$ is the
redshift of the background quasar. 

Similarly, for a line of sight in the control sample, the net RM ($\rmc$) in
the observer's frame is given by,
\begin{equation}
\rmc = \frac{\rm RM_{qso,c}}{(1+z_{\rm qso, c})^2} + {\rm RM_{IGM,c} + RM_{MW,c} + \delta_{\rm RM,c}},
\end{equation}
where the subscript `c' refers to the control sample.

If quasars in the target and the control samples are chosen such that $\rm
RM_{qso}^\prime$ and $\rm RM_{qso,c}$ have the same statistical properties,
then statistical comparison between the distributions of $\rmt$ and $\rmc$ can
yield the excess contribution from ${\rm RM_{gal}}\,{(1+z_{\rm gal})^{-2}}$.
Most of the previous studies where this method is used have attributed
statistical differences to turbulent magnetic fields, with the field strengths
computed assuming Gaussian statistics for $\rmt$ and $\rmc$
\citep[e.g.,][]{berne08, berne13}.

Here we investigate the effects of large-scale magnetic fields in the
intervening galaxies on the distribution of RM. First, we derive the
probability distribution function (PDF) of $\rmgal$ analytically for a single
galaxy. Then, we consider the statistical distribution of $\rmgal$ for an
ensemble of galaxies. A detailed treatment can be found in \citet{basu18}.

\section{Magneto-ionic medium in intervening galaxy population}

To derive the PDF of $\rmgal$ analytically, we have adopted a simple set of
assumptions for the magneto-ionic medium in the intervening galaxies: (i) the
large-scale magnetic fields in each galactic disc have axisymmetric geometry
with magnetic pitch angle $p$, (ii) both the large-scale magnetic field
strength ($B$) and the free electron density ($n_{\rm e}$) decrease
exponentially with radius $r$, (iii) the magnitude and geometry of both
magnetic field and electron density do not vary with height from the mid-plane,
(iv) the RM contributed by turbulent magnetic fields in the intervening
galaxies is insignificant within the three-dimensional illumination beam
passing through the galaxies, and (v) the RM contributed by the background
quasar has been statistically isolated from $\rmt$. 

Under these assumptions, the RM for each line of sight is given as
\citep[see][]{berkh97, basu18},
\begin{equation}
\rmgal = -0.81\, n_0\, B_0\, {\rm e}^{-r/r_0^\prime}\,\cos (\theta - p) \, h_{\mathrm{ion}} \tan i.
\end{equation}
Here, $n_0$ and $B_0$ are the free electron density and the strength of the
axisymmetric large-scale magnetic field at the galactic centre, $\theta$
is the azimuthal angle, $h_{\rm ion}$ is the thickness of the ionized disc,
$i$ is the inclination angle of the galactic disc with respect to the plane of
the sky and $r_0^\prime$ is the radial scale-length of the product $n_{\rm
e}(r)\,B(r)$. 

For an ensemble of intervening galaxies, the quasar sightlines can intersect
each galaxy at any radius, inclination angle, and azimuthal angle. For galaxies
in the target sample, we assume the galactic discs to be inclined uniformly
with $i$ in the range 0 to $90^\circ$. The angle $\theta$ is distributed
uniformly between 0 and $360^\circ$. Since it is plausible that the currently
available data base of absorption line samples is incomplete in terms of range
of radii probed by the background quasars \citep{basu18}, we assume that the
impact radii are distributed uniformly in the range $R_{\rm min}$ to $R_{\rm
max}$.

\begin{figure*}
\begin{center}
\begin{tabular}{cc}
 {\mbox{\includegraphics[height=2.4in]{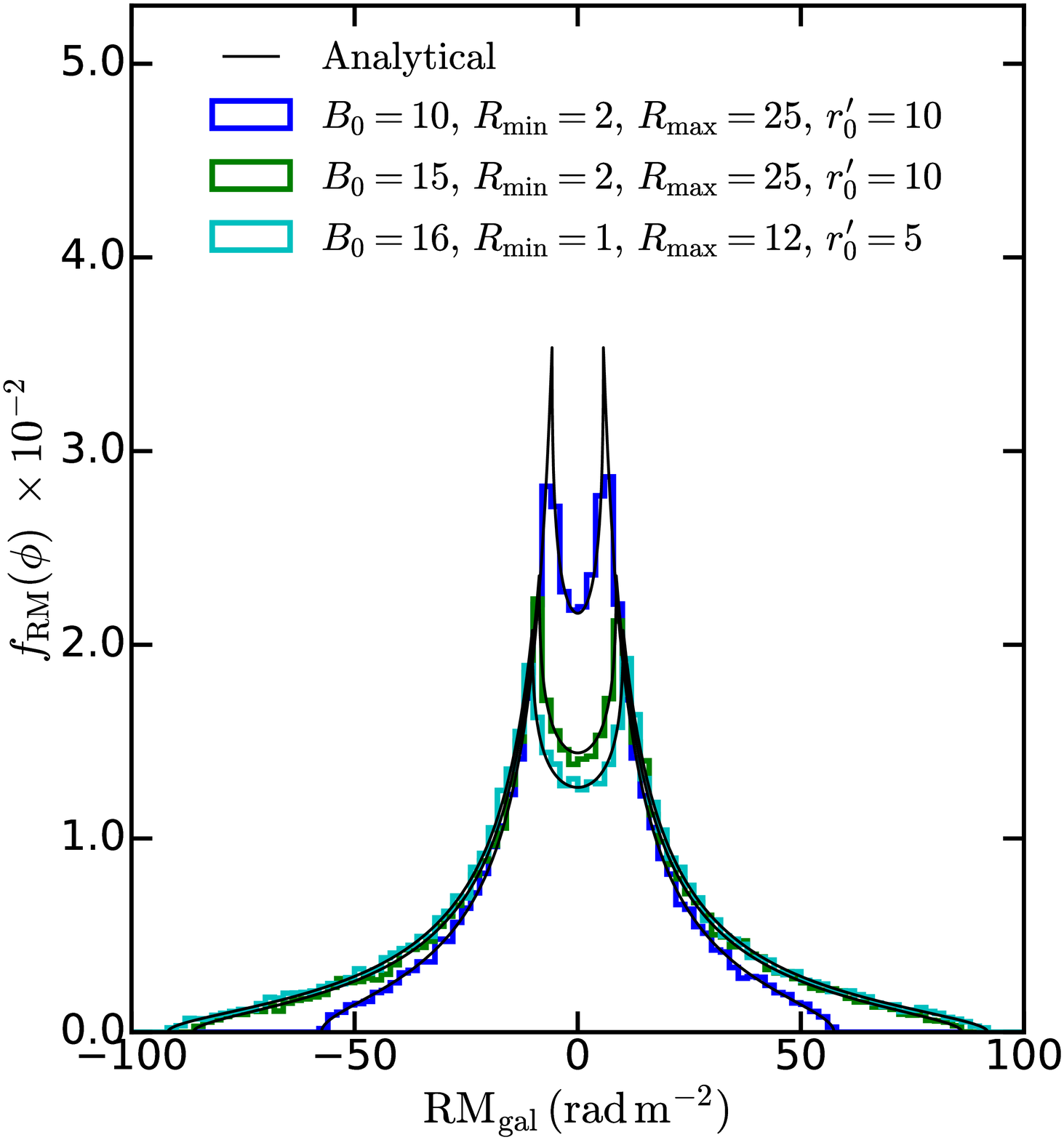}}} &
 {\mbox{\includegraphics[height=2.6in]{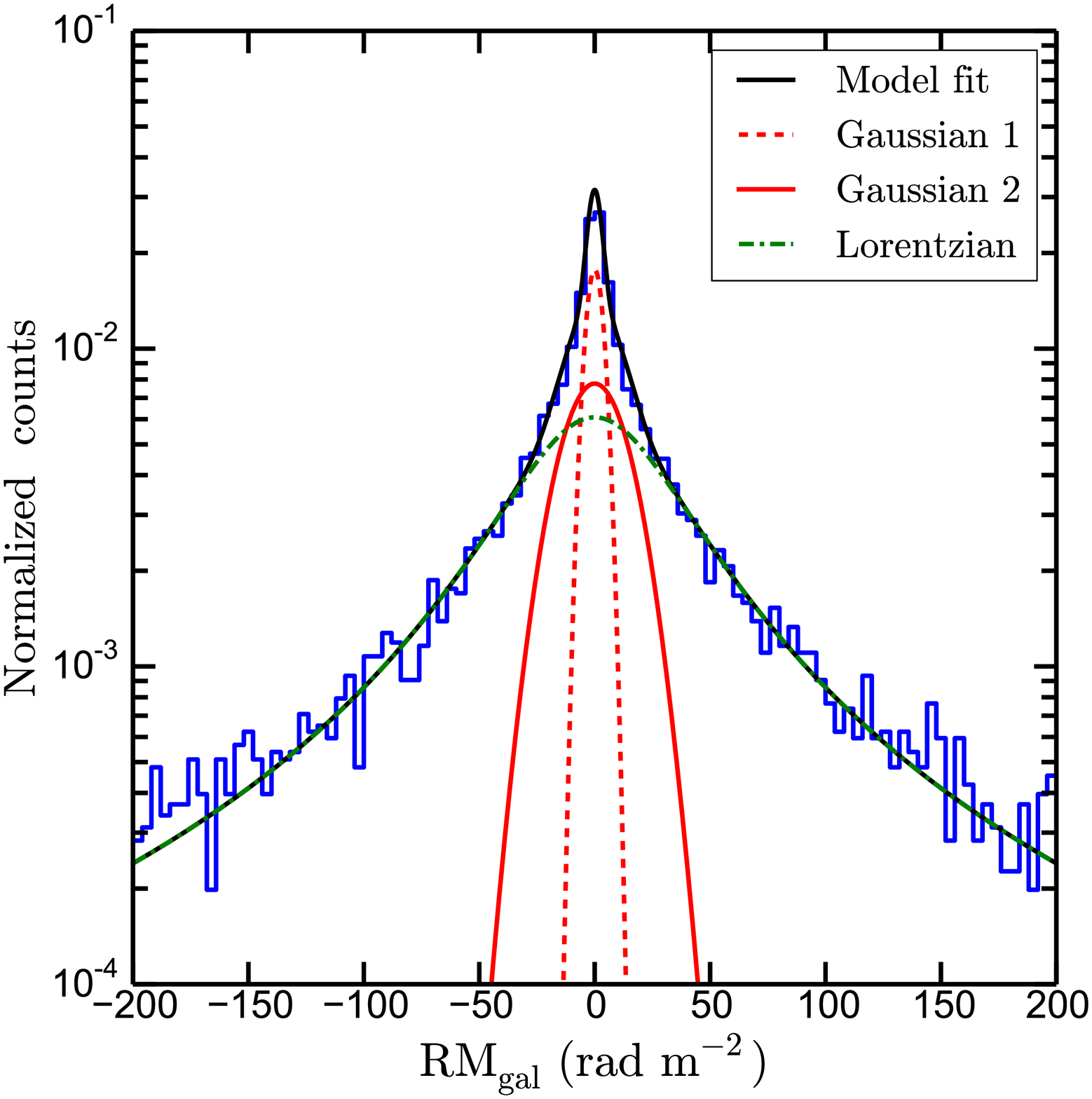}}}\\ 
\end{tabular}
\end{center}
 \caption{{\it Left}: Distribution of $\rmgal$ for a single galaxy inclined at
$30^\circ$. The black curves are the analytical PDF given by
Eq.~\eqref{eq:pdf_rm}. {\it Right:} Distribution of $\rmgal$ for a sample of
10\,000 galaxies. $B_0$ and $n_0$ of the population have Gaussian distributions
with $\langle B_0\rangle = 15\,\mu$G, $\langle n_0\rangle = 0.03\,\rm cm^{-3}$,
$\sigma_{B_0}=5\,\mu$G and $\sigma_{n_{\rm e}} = 0.01\,\rm cm^{-3}$. The
distribution is modelled as a sum of one Lorentzian and two Gaussians, which
are shown as the different lines.}
   \label{fig1}
\end{figure*}

\section{PDF of RM of intervening galaxy population}

Under the assumptions mentioned above, the PDF of $\rmgal$, $f_{\rm RM}(\phi)$,
for a single galaxy has the following analytical form:
\begin{equation} 
f_{\rm RM}(\phi) = 
\begin{cases}
~\frac{r_0^\prime}{\pi\,(R_{\rm max} - R_{\rm min})}\,\frac{1}{|\phi|}\left[ \arcsin \left(\frac{|\phi|}{a^\prime} \right)-  \arcsin\left(\frac{|\phi|}{b^\prime}\right) \right], &  -a^\prime \le \phi \le a^\prime,
\\
~\frac{r_0^\prime}{\pi\,(R_{\rm max} - R_{\rm min})}\,\frac{1}{|\phi|}\,\arccos \left(\frac{|\phi|}{b^\prime} \right), & \phi \in [-b^\prime,-a^\prime) \cup (a^\prime,b^\prime].
\\
\end{cases}
\label{eq:pdf_rm}
\end{equation}
Here, $a^\prime = 0.81\,n_0\,B_0\,h_{\rm ion} \tan i\, {\rm e}^{-R_{\rm
max}/r^\prime_0}$ and $b^\prime = 0.81\,n_0\,B_0\,h_{\rm ion} \tan i\, {\rm
e}^{-R_{\rm min}/r^\prime_0}$. The left-hand panel of Fig.~\ref{fig1} shows the
analytical PDF of $\rmgal$ for a single galaxy with $i=30^\circ$ (black curve)
and, for comparison, the simulated distribution for 10\,000 lines of sight as
histograms.

\begin{figure}[t]
\begin{center}
 \includegraphics[width=2.5in]{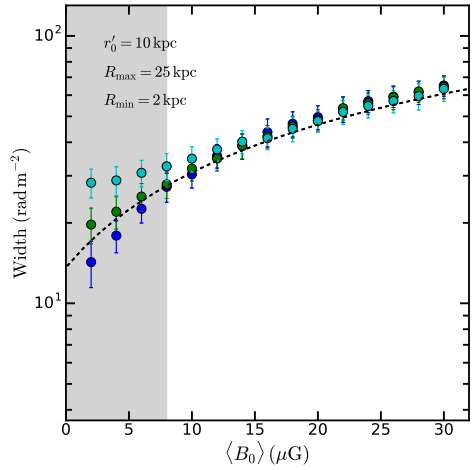} 
\end{center}
 \caption{Variation of the width of the Lorentzian component of the
distribution of $\rmgal$ in the target sample as a function of $\langle
B_0\rangle$. Blue, green and cyan symbols are for $\sigma_{B_0}=2,\,5\, {\rm
and}\,10\,\mu$G. The dashed line is the best-fit model given by
Eq.~\eqref{eq:width}. The shaded region roughly represents the region where
$\langle B_0\rangle \lesssim \sigma_{B_0}$ and the points within were excluded
from the fitting.}
   \label{fig2}
\end{figure}

The right-hand panel of Fig.~\ref{fig1} shows a simulated PDF of $\rmgal$ for a
sample of 10\,000 intervening disc galaxies, where we have assumed Gaussian
distributions for $B_0$ (with mean $\langle B_0\rangle$ and standard deviation
$\sigma_{B_0}$) and $n_0$ (with mean $\langle n_0 \rangle$ and standard
deviation $\sigma_{n_{\rm e}}$) within the population. We find that the
distribution of $\rmgal$ can be accurately approximated by the sum of one
Lorentzian and two Gaussian components as shown in the right panel of
Fig.~\ref{fig1}. From Eq.~\eqref{eq:pdf_rm} and Fig.~\ref{fig1} (left panel),
we find that the width of the PDF of $\rmgal$ for a single galaxy depends on
the parameters $a^\prime$ and $b^\prime$ and, therefore, on $B_0$. Thus, we
expect that the width $w$ of the Lorentzian component to be related to $\langle
B_0\rangle$ for a sample of intervening galaxies. In Fig.~\ref{fig2}, we show
the variation of $w$ with $\langle B_0\rangle$ for various choices of
$\sigma_{B_0}$. In the regime $B_0 \gtrsim \sigma_{B_0}$ (the un-shaded area in
Fig.~\ref{fig2}), the variation of $w$ is approximated by,
\begin{equation}
w(\langle B_0\rangle) = 13.6 + 1.8\,\langle B_0\rangle - 0.0076\,\langle B_0\rangle^2,
\label{eq:width}
\end{equation}
with $\langle B_0\rangle$ in $\mu$G and $w$ in rad\,m$^{-2}$, and is shown as the
dashed line in Fig.~\ref{fig2}. Thus, $w$ can be used to determine the mean
strength of the large-scale magnetic field in a sample of intervening galaxies.
 
\section{Conclusion}

Our study suggests that the distribution of $\rmgal$ due to axisymmetric
large-scale magnetic fields in intervening disc galaxies is non-Gaussian and
that the distribution can be empirically modelled as the sum of one Lorentzian
and two Gaussian components. The width of the Lorentzian component can be used
to estimate the mean large-scale magnetic field strength in a population of
intervening galaxies using Eq.~\eqref{eq:width} derived for typical values of
the physical parameters, such as, $h_{\rm ion} = 500$ pc, $R_{\rm min}=2$ kpc,
$R_{\rm max}= 25$ kpc and $r_0^\prime = 10$ kpc. The $\langle B_0\rangle$
estimated using Eq.~\eqref{eq:width} lies within $\sim50\%$ of the true
value in the absence of additional constraints on the physical parameters.

\end{document}